\documentclass[prl,twocolumn,amsmath,showpacs]{revtex4}
\input {epsf.sty}
\usepackage{epsfig}
\usepackage{graphicx}
\usepackage{amsmath,amsthm,amssymb,latexsym,amsfonts,times}
\usepackage{bm}

\usepackage{epstopdf}
\begin{document}
\title{Anomalous Attenuation of Transverse Sound in $^3$He}
\author{J.P. Davis, J. Pollanen, H. Choi, J.A. Sauls and W.P. Halperin}
\affiliation{Department of Physics and Astronomy, Northwestern University,
Evanston, Illinois 60208}
\author{A.B. Vorontsov}
\affiliation{Department of Physics, University of Wisconsin-Madison,  Madison,
Wisconsin 53706}

\date{Version \today}


\begin{abstract}  We present the first measurements of the attenuation of
transverse sound in superfluid $^3$He-B.  We use fixed path length 
interferometry
combined with the magneto-acoustic Faraday effect to vary the effective path
length by a factor of two, resulting in absolute values of the attenuation.  We
find that attenuation is significantly larger than expected from the 
theoretical
dispersion relation, in contrast to the phase velocity of transverse 
sound.   We
suggest that the anomalous attenuation can be explained by surface Andreev
bound states.
\end{abstract}

\maketitle

\vspace{11pt}

	Over fifty years ago Landau published his seminal works on 
Fermi liquid theory
\cite{Lan56, Lan57}.  In the second of these works, on the collective 
dynamics of
Fermi liquids, he predicted that there would be collisionless sound 
modes outside
the hydrodynamic limit, called zero sound.  The crossover from 
hydrodynamic sound
to longitudinal zero sound was discovered in the normal state of 
$^3$He in 1966 by
Abel, Anderson and Wheatley \cite{Abe66}.  Along with longitudinal zero sound,
Landau predicted that for certain values of the Fermi liquid interaction
parameters
\cite{Bay78} there should be a collisionless collective mode called transverse
zero sound.  The constraint on the Fermi liquid interaction parameters is
essentially that the transverse sound velocity, $c_t$, be greater 
than the Fermi
velocity, $v_F$, otherwise the transverse wave can decay into incoherent
quasiparticles in a process called Landau damping
\cite{Lea76}.  This condition is likely satisfied over the entire 
range of liquid
$^3$He \cite{Hal90}. However, attempts to observe transverse sound (TS) in the
normal state of $^3$He \cite{Roa76} have proven unsuccessful 
\cite{Flo76}, due in
part to high attenuation.

\begin{figure}[b]
\centerline{\includegraphics[width=2.9in]{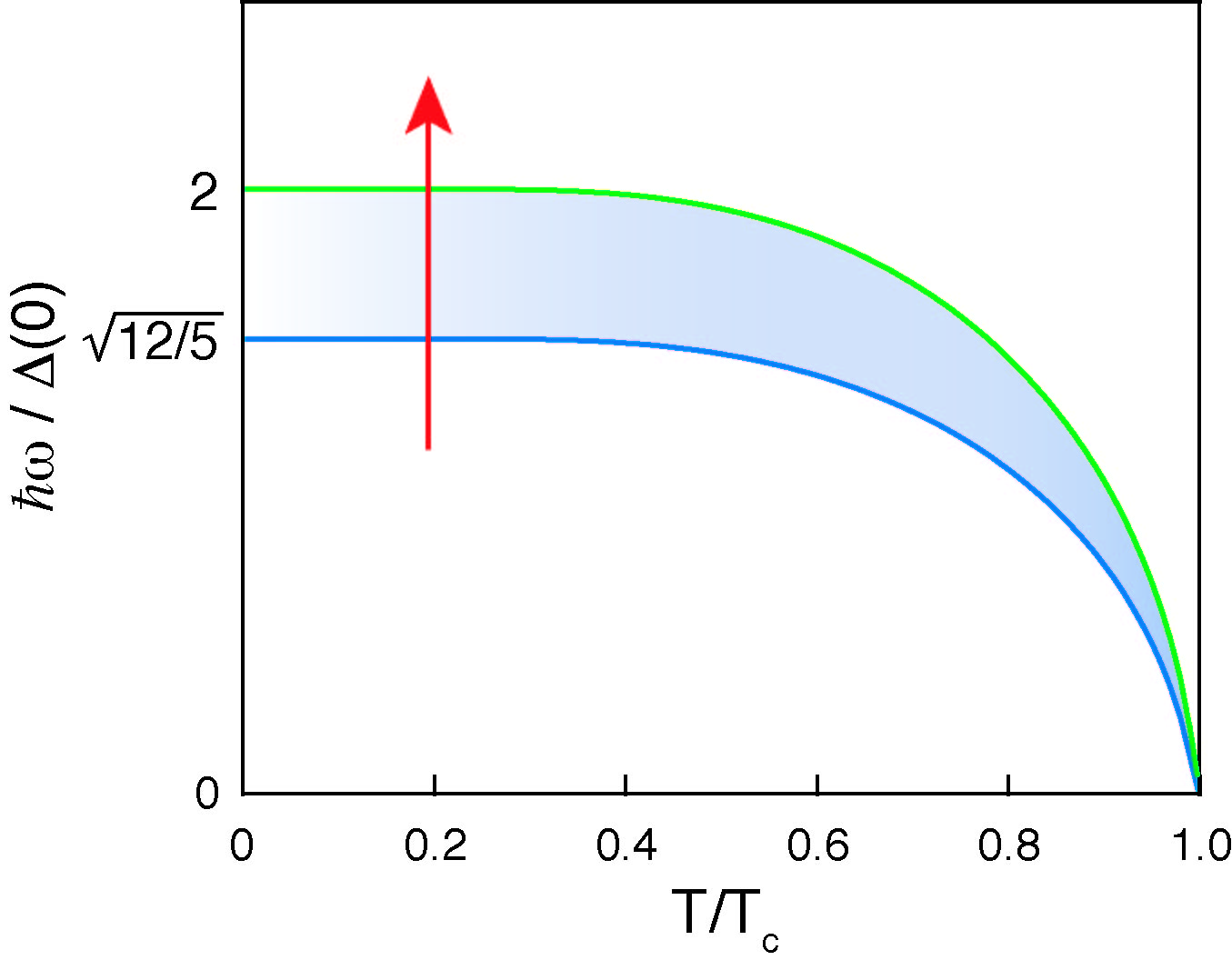}}
\caption{\label{fig1} (color online) The energy of pair-breaking (green curve)
and the ISQ-mode (blue curve) as a function of temperature normalized 
to $T_c$.  TS
propagates only in the shaded blue region. The low temperature pressure sweep
technique follows a path represented by the red arrow.}
\end{figure}

Predictions for the fate of TS in the superfluid state of $^3$He were 
pessimistic,
since the number of unpaired quasiparticles decreases as the energy 
gap opens up
\cite{Leg66, Mak74}.  In 1993, Moores and Sauls (MS) \cite{Moo93} showed that
these ideas were incomplete and instead TS would be enhanced in the B-phase of
superfluid $^3$He due to the off-resonant coupling of transverse currents to an
order parameter collective mode, called the imaginary squashing mode 
(ISQ).  They
showed that the dispersion relation for TS, in the long wavelength limit, was
given by:
\begin{equation}\label{dispersion}
        \frac{\omega^2}{q^2 v_{F}^2} = \Lambda_{0} +
\Lambda_{2^{-}}\frac{\omega^{2}}{(\omega +
i\Gamma)^{2}-\Omega_{2^{-}}^{2}-\frac{2}{5}q^{2}v_{F}^{2}},
\end{equation} where $q$ is the complex wavevector, $q = k + i\alpha$,
$k$ is the real wavevector, $\alpha$ is the attenuation, and the 
phase velocity is
$c_t =
\omega/k$.  The ISQ-mode frequency closely follows the temperature and pressure
dependence of the energy gap, $\Delta(T,P)$,
$\Omega_{2^{-}}(T,P) = a_{2^{-}}(T,P)\Delta(T,P)$, where $a_{2^{-}}
\approx \sqrt{12/5}$
\cite{Moo93,Dav06,Dav08} and the ISQ-mode width is given by $\Gamma$
\cite{Moo93,Ein84}. It is customary to label this mode $2^{-}$, 
according to its
total angular momentum quantum number and its parity under particle-hole
conversion. The first term on the right hand side of 
Eq.~\ref{dispersion} is the
quasiparticle  background, the contribution to the dispersion in the absence of
coupling to the ISQ-mode, and the second term gives the off-resonant coupling
strength to the ISQ-mode.  This off-resonant coupling produces a dramatic
increase in the phase velocity of TS near the mode \cite{Dav08b}, 
lifting it well
above the Fermi velocity and thereby reducing Landau damping.  But this is only
allowed above the ISQ-mode energy and below the pair-breaking energy, 
shown as the
blue shaded region in Fig.~\ref{fig1}.

	The predictions of MS \cite{Moo93} prompted a new exploration 
for TS in $^3$He,
which yielded more fruitful results than in the normal state
\cite{Kal93,Dav06,Dav08,Dav08b,Lee99,Dav08c,Dav08d}. Additionally, MS 
predicted a
magneto-acoustic Faraday effect (AFE). Its observation by Lee {\it et al.}
\cite{Lee99} confirmed the existence of TS in liquid $^3$He, the only 
liquid where
transverse sound is known to propagate.  In this Letter, we present 
measurements
of the absolute attenuation of transverse sound in superfluid $^3$He-B.  These
measurements show a larger attenuation than expected from Eq.~\ref{dispersion},
which we suggest arises from surface Andreev bound states.

To measure the attenuation, an acoustic cavity was constructed with 
one wall as a
shear transducer ($AC$-cut) and the other as an optically polished quartz
reflector. The thickness of the acoustic cavity is $D = 
31.6~\pm~0.1~\mu$m and was
filled with liquid $^3$He.  This is small enough that standing waves of TS are
able to form in the cavity \cite{Dav08,Dav08b,Dav08d}.  Here we use 
the $13^{th}$
to the $25^{th}$ transducer harmonics (76 to 147 MHz).  We note that the TS
velocity is a sensitive local indicator of the temperature in the 
acoustic cavity,
which was used to ensure that there was no heating from the transducer or other
sources.  Furthermore, the acoustic cavity walls are guaranteed parallel via a
spring loaded set-up that maintains the cavity spacing at all times and
temperatures.  Information on these experimental techniques have been described
in detail elsewhere
\cite{Dav06,Dav08,Dav08b,Dav08c}.  Throughout we use the 
weak-coupling-plus (WCP)
gap of Ref.~\cite{Rai76}, tabulated in Ref.~\cite{Hal90} with values of
$T_c$ given by Greywall \cite{Gre86}.  And in $\Lambda_{0}$ and
$\Lambda_{2^{-}}$ we use the Tsuneto function calculated using the 
WCP gap and all
Fermi liquid parameters up to $l\leq2$ \cite{Dav08b,Dav08c,Dav08d}.

The electrical impedance of the shear transducer is sensitive to the 
standing TS
wave at the surface of the transducer and was monitored with a continuous wave
impedance bridge \cite{Dav08c}.  The output of the bridge is,
\begin{equation}\label{VZ} V_Z~=a+b\cos\theta\sin\bigg(\frac{2D\omega}{c_t} +
\phi \bigg),
\end{equation}	  where $\theta$ is the angle of the polarization of 
the TS wave at
the surface of the transducer relative to the intrinsic polarization 
of the shear
transducer, $c_t$ is the phase velocity of TS and $\phi$ is a fixed phase that
depends on the experimental conditions.  A smoothly varying 
background of acoustic
impedance \cite{Nag07} is represented by $a$ and the attenuation, $\alpha$ is
proportional to -ln\,$b$.  By varying the temperature or pressure at fixed
acoustic frequency we sweep ${\hbar
\omega}/{\Delta(T,P)}$ (see Fig.1), changing the acoustic frequency relative to
the energy of the ISQ-mode and therefore
$c_t$, producing oscillations in $V_Z~$\cite{Dav08,Dav08b}.

\begin{figure}[t]
\centerline{\includegraphics[width=3.3in]{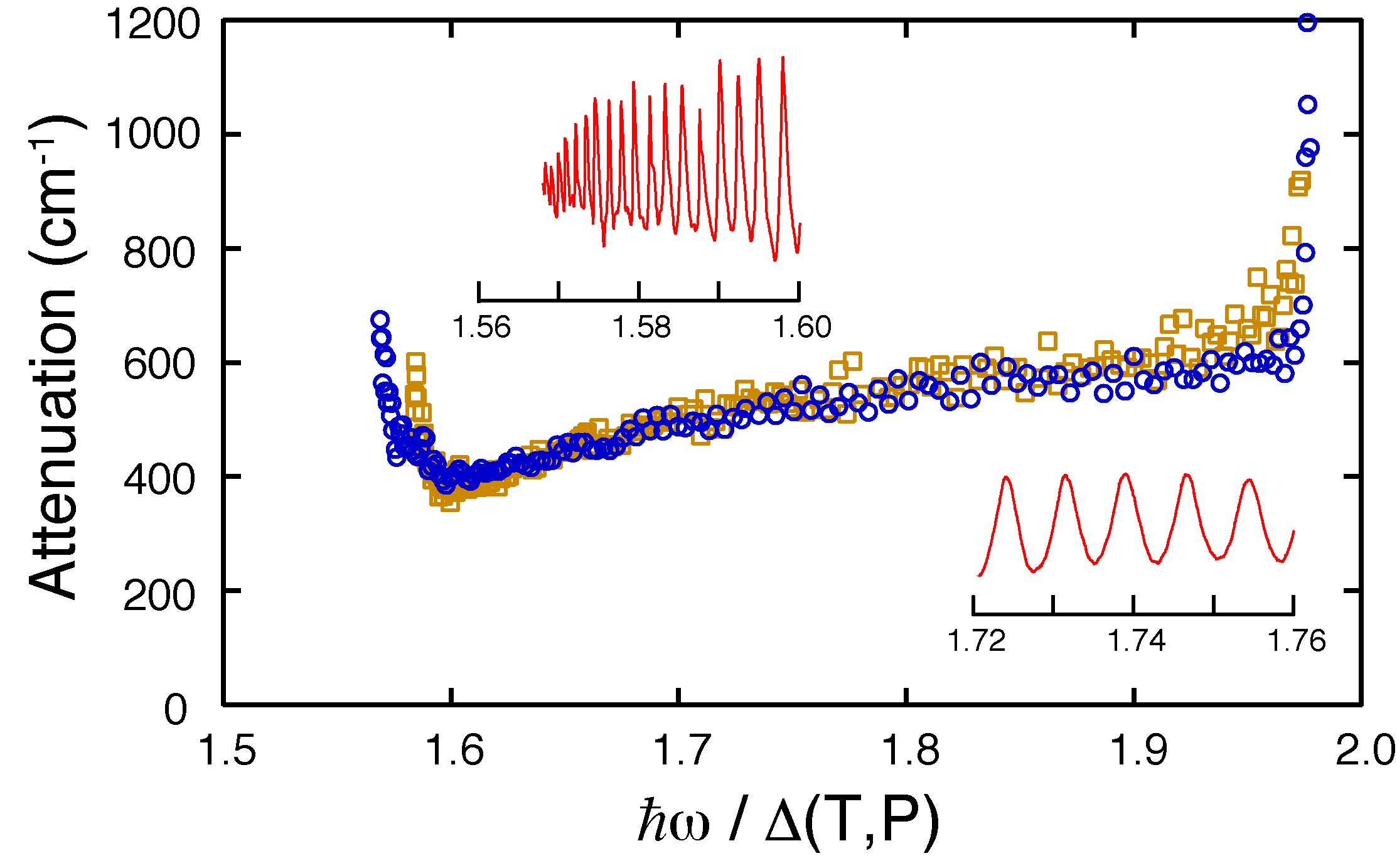}}
\caption{\label{fig2} (color online) Attenuation of transverse sound as a
function of energy normalized to the energy gap at constant  temperature,
$\approx 550~\mu$K from a pressure sweep.  Blue circles and gold squares
are for 88 and 111.5 MHz respectively.  In the insets we show examples
of the acoustic interference oscillations in $V_Z$ over  small energy ranges for comparison,
both on the same energy and amplitude scales. }
\end{figure}

	In previous reports \cite{Dav06, Dav08, Dav08b} we noted that 
TS attenuation is
inversely related to the amplitude of the acoustic response 
oscillations, but we
could not make a quantitative interpretation.  Here, on the other 
hand, we obtain
the absolute value of the TS attenuation, taking advantage of the 
acoustic Faraday
effect.  Using the AFE, we rotate the linear polarization of the TS 
waves in the
acoustic cavity
\cite{Dav08} by applying a magnetic field along the sound propagation 
direction.
When the polarization is rotated by
$\pi/2$, there is a minimum in the envelope of acoustic response oscillations,
modeled by the $\cos\theta$ in Eq.~\ref{VZ}.  The angle $\theta$ is 
proportional
to the path length and consequently, at this minimum, the standing 
waves to which
our transducer is sensitive have an effective path length of $4D$.  Under these
conditions smaller amplitude oscillations occur twice as frequently in the same
interval of a temperature or pressure sweep.  Comparing the amplitude 
of the waves
with a path  length of $2D$ to the amplitude of the waves with a path 
of $4D$ we
find the absolute attenuation,
$\alpha$, for one particular frequency, temperature and pressure:
\begin{equation}
\alpha = - \frac{1}{2D}\ln \bigg[\frac{b_{4D}}{b_{2D}}\bigg].
\end{equation}  The absolute value of the attenuation at all temperatures and
pressures can then be determined, as shown in Fig.~\ref{fig2}, with 
\emph{no} fit
parameters.  The increased attenuation at the low energy end of 
Fig.~\ref{fig2} is
from the ISQ-mode and the increase in the attenuation near $2\Delta$ originates
from the
$2\Delta$-mode, recently reported
\cite{Dav08b}.  The ISQ-mode frequency has a weak pressure dependent deviation
from $\sqrt{12/5}\Delta(T,P)$
\cite{Dav06,Dav08} which is reflected in the offset of the up-turns 
in attenuation
at low energy for the two frequencies in Fig.~\ref{fig2}.  With our 
technique we
are able to observe propagating TS with an attenuation as high as 
1000 cm$^{-1}$.
As yet, we have not found any indication of propagating TS in the 
normal state of
$^3$He and its observation will require overcoming this higher than expected
attenuation
\cite{Cor69}.  We note in passing that our measurements were performed in the
quantum limit of attenuation described by Landau \cite{Lan57}, with 
$\hbar \omega
/ 2\pi k_B T = 1.2$ (1.3) for the 88 (111.5) MHz data.

\begin{figure}[t]
\centerline{\includegraphics[width=3.4in]{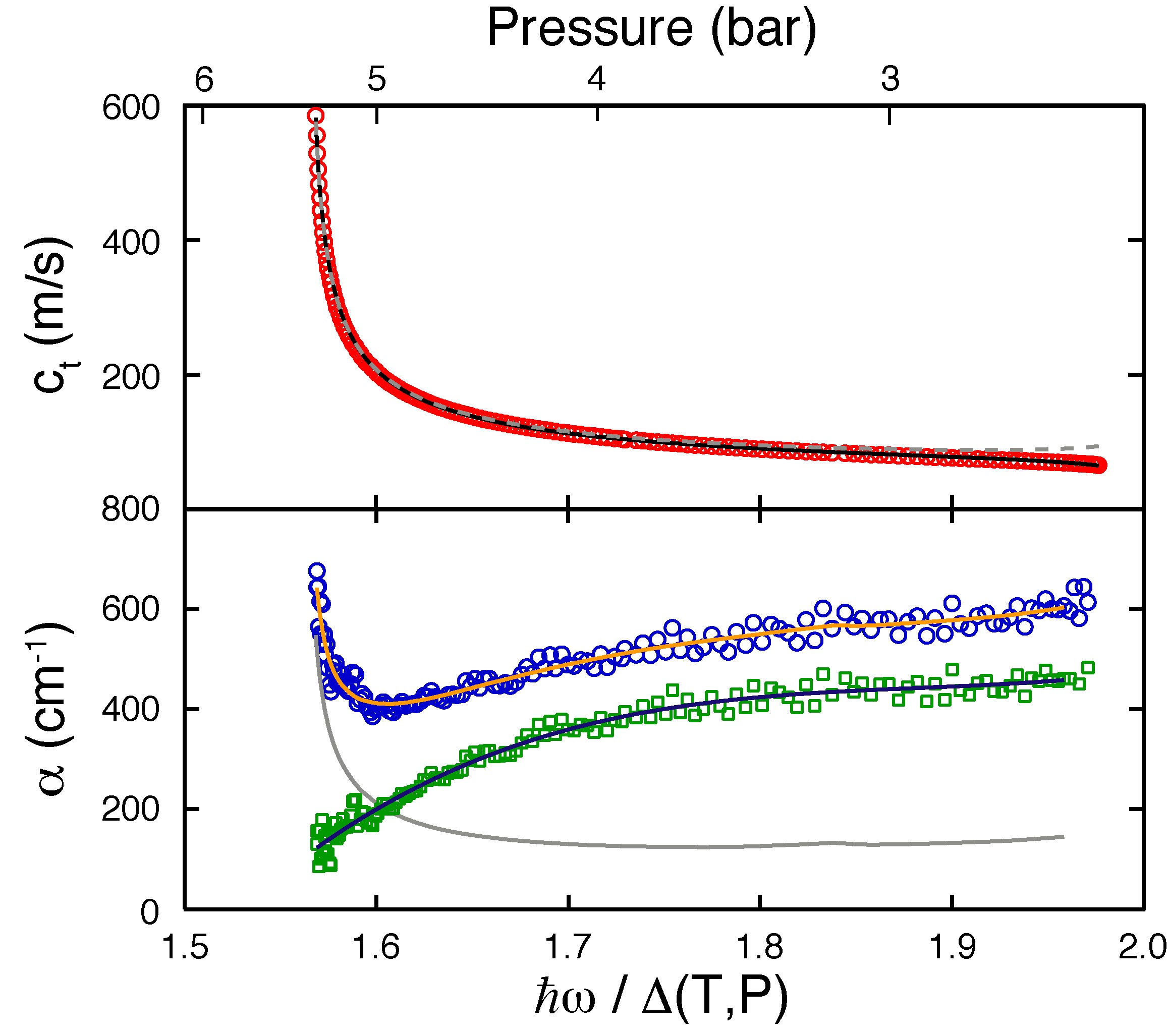}}
\caption{\label{DeconvolvedAtten} (color online) Phase velocity of transverse
sound (upper panel) and attenuation (lower panel) at $88$ MHz at
$\approx550~\mu$K in zero magnetic field.  In the upper panel, the 
red circles are
the data, the dashed grey curve is calculated from 
Eq.~\ref{dispersion}, and the
black curve is the dispersion that accounts for the $2\Delta$-mode
\cite{Dav08b}.  In the lower panel, the attenuation data (blue 
circles) has been
deconvolved by subtracting the contribution from coupling to the ISQ-mode (grey
curve) calculated from Eq.~\ref{dispersion} leaving an anomalous attenuation
given by the green squares.}
\end{figure}

\begin{figure}[t]
\centerline{\includegraphics[width=3.3in]{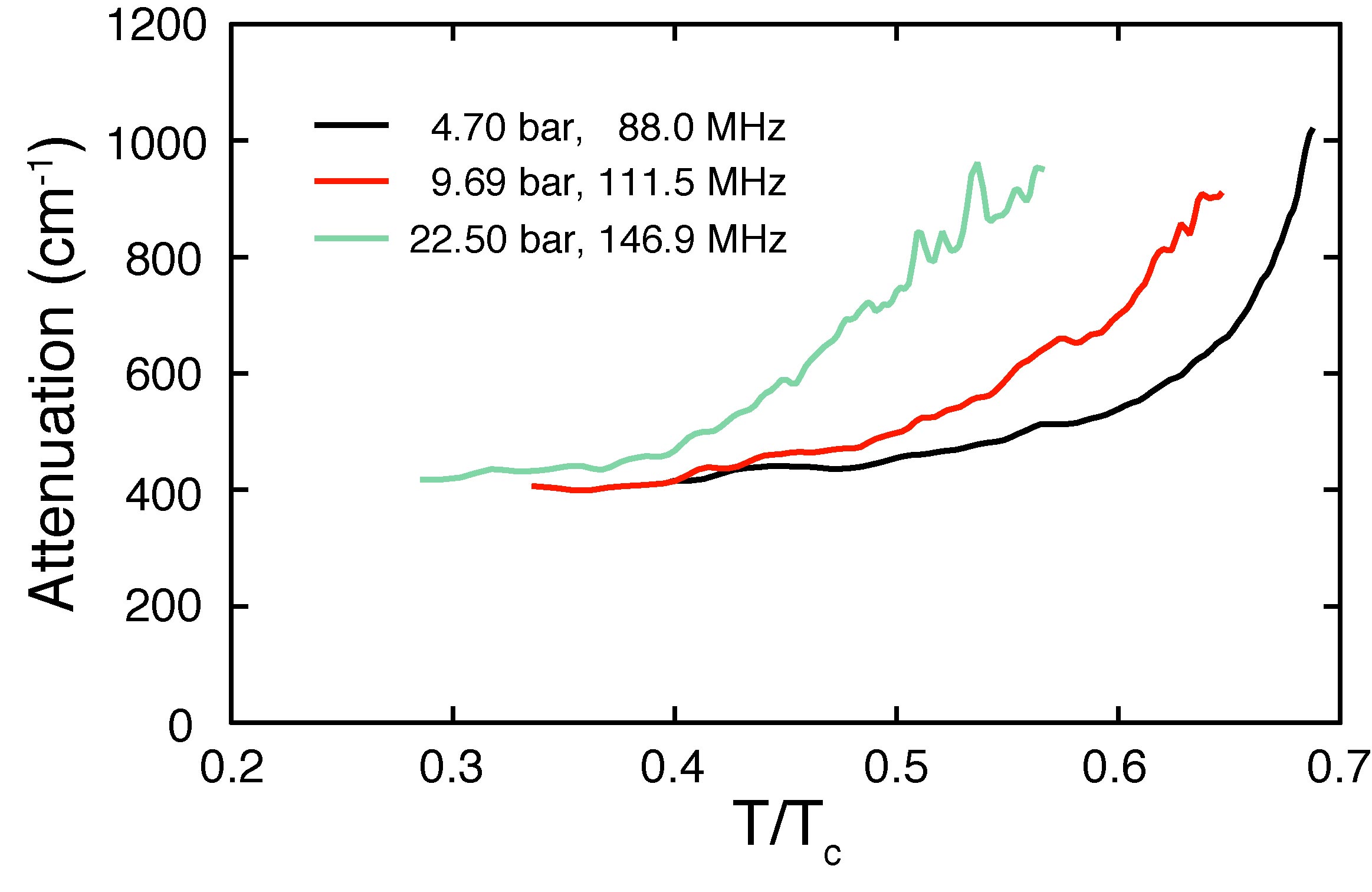}}
\caption{\label{AttenTempDep} (color online) The temperature 
dependence of the TS
attenuation.  The pressures are chosen for the acoustic frequencies such that
$\hbar\omega/\Delta$ is within the shaded region of Fig.~\ref{fig1} and never
crosses the ISQ-mode.  The data  has been minimally smoothed and is shown as a
line for clarity. Quasiparticle-quasiparticle scattering, seen as an increasing
attenuation at high
$T/T_c$, is expected to decrease to zero at zero temperature. 
Instead, there is a
crossover from the quasiparticle dominated region at high $T/T_c$ to 
a temperature
independent anomalous region at low
$T/T_c$.  The low temperature end-points of the data correspond to
$\hbar\omega/\Delta \approx$
   1.64.}
\end{figure}

The contribution to the attenuation from the ISQ-mode can be calculated from
Eq.~\ref{dispersion} with only a single fit parameter: the width of the
ISQ-mode.  We use the form
$\Gamma = \Gamma_{c}e^{-\Delta/k_{B}T}$, where $\Gamma_{c} =
\Gamma_{0} T_{c}^2$, and $\Gamma_{0}$ is pressure independent.  The ISQ
attenuation is shown separately, for the 88 MHz data, by the grey curve in
Fig.~\ref{DeconvolvedAtten}. In order to represent  the observed non-monotonic
dependence of attenuation on
$\hbar \omega/\Delta$ it is clear that there must be an additional 
contribution.
This unexpected behavior apparently increases smoothly with energy and then
saturates, $\hbar \omega /\Delta \approx 1.7$. To obtain a quantitative
assessment of this anomalous attenuation we must choose a value for 
$\Gamma_{0}$
which, if taken either too large or too small, will introduce an unphysical,
sharp kink at $\hbar \omega /\Delta \sim 1.6$.  Our final result using
$\Gamma_{0} = 9.5 \pm 2$ MHz/mK$^2$ is given by the green squares in
Fig.~\ref{DeconvolvedAtten}. Since the ISQ-mode attenuation dominates only near
the mode the subtracted result is largely unaffected by our choice of
$\Gamma_{0}$, which we find to be a factor of three larger than previously
suggested~\cite{Ein84}, based on a less accurate measurement of the ISQ-mode
width \cite{Hal82}. Additionally, we find that the
anomalous attenuation approaches the temperature independent value at  low
temperatures given in Fig.~\ref{DeconvolvedAtten}, as demonstrated by 
temperature
sweeps in Fig.~\ref{AttenTempDep}.

\begin{figure}[t]
\centerline{\includegraphics[width=3.3in]{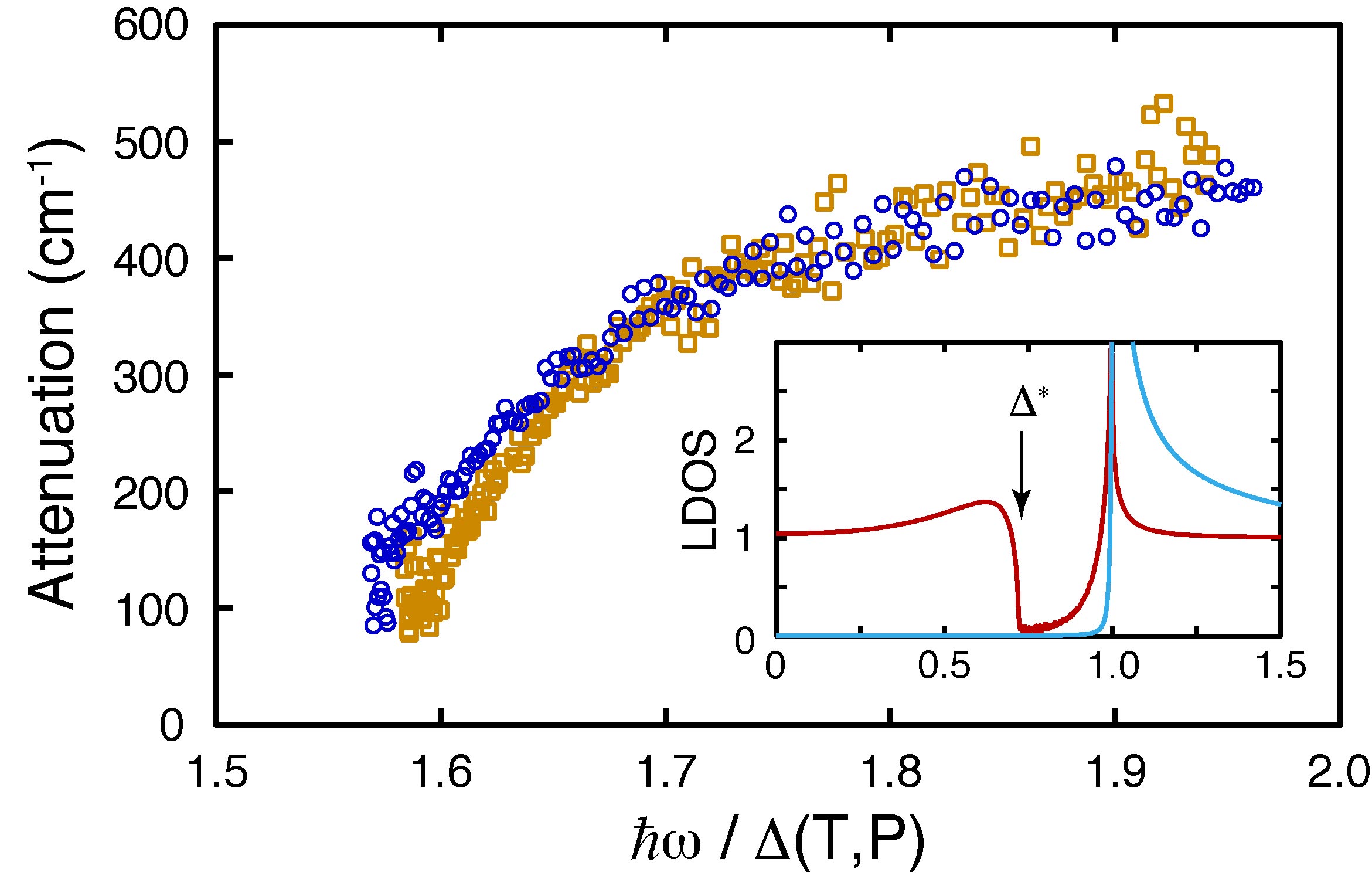}}
\caption{\label{ExtraAttenBoth} (color online) The anomalous 
attenuation, for 88
MHz (blue circles) and 111.5 MHz (gold squares).  A smooth crossover in the
attenuation appears at
$\hbar\omega = \Delta+\Delta^{\star} \approx 1.7\Delta$.  The inset shows the local
density of states, as a function energy normalized to $\Delta$, at 
the transducer
surface (red) and in the bulk $^3$He (light blue) at $T/T_c=0.5$ for diffusive
boundary conditions.}
\end{figure}

In contrast to the attenuation, we have found that the phase velocity of TS is
accurately accounted for by the dispersion relation for the order parameter
collective mode, Eq.~\ref{dispersion},
\cite{Dav08b,Dav08d}, as shown in the upper panel of
Fig.~\ref{DeconvolvedAtten}.  We infer that the anomalous attenuation cannot be
associated with order parameter collective modes. Furthermore, the 
data at 88 and
111.5 MHz are nearly identical, shown in Fig.~\ref{ExtraAttenBoth}, indicating
that the attenuation is  not explicitly dependent on frequency at the 
same values
of
$\hbar \omega/\Delta$, nor does it depend on temperature in the low temperature
limit, Fig.~\ref{AttenTempDep}.  On this basis we can rule out
quasiparticle-quasiparticle scattering  as the source, since this 
mechanism should
decrease to zero exponentially at low temperatures.  We have applied magnetic
fields up to 300 G along the TS propagation direction and have found that the
attenuation does not depend on magnetic field, outside of the regions of field
induced birefringence from order parameter collective modes (AFE).  We suggest
that the anomalous attenuation might be attributed to the interaction 
of TS waves
with surface Andreev bound states (SABS).

SABS play an important role in the understanding of unconventional
superconductors and superfluids.  For example, SABS have been studied in
tunneling experiments in Sr$_2$RuO$_4$ \cite{Lau00} and the high $T_c$
superconductors \cite{Cov97,Kru99}.  In superfluid
$^3$He they have been found to dominate the transverse acoustic impedance
\cite{Aok05} and have been observed in the surface specific heat
\cite{Cho06}.  Moreover, in the absence of excited quasiparticles, there is no
coupling between a transverse
    transducer and $^3$He, for example when the scattering at the transducer
surface is specular
    \cite{Moo93b}.  However, quasiparticles that scatter diffusely transfer
     momentum parallel to the transducer surface and couple to transverse
currents in the $^3$He \cite{Moo93b}.  These local excitations are the bound
states (SABS).  In $^3$He-B they have a characteristic energy given by
$\Delta^{\star}$, \cite{Nag07,Vor03} the upper limit of the density of states
band which we show integrated over all trajectories in the
inset of Fig.~\ref{ExtraAttenBoth} (red trace). These midgap states are
responsible for structure observed in the temperature dependence of 
the acoustic
impedance
\cite{Aok05,Nag07} between the transducer and helium and should also affect the
amplitude of a transverse sound wave reflected from a surface.  Excitation of
SABS will attenuate the wave and we expect this to follow the frequency
dependence of the imaginary part of the acoustic impedance \cite{Nag07},
increasing with frequency up to
$\hbar\omega =
\Delta+\Delta^{\star}$ and then leveling off.  This scenario is qualitatively
consistent with the  attenuation shown in Fig.~\ref{ExtraAttenBoth}
where we observe a smooth but distinct crossover near
$\hbar\omega \approx 1.7\Delta$ to a regime of anomalous 
attenuation at higher
energy.  With this interpretation our results are in good agreement with the
theoretical value for
$\hbar\omega =
\Delta+\Delta^{\star} = 1.75\Delta$,  at
$T/T_c \sim 0.4$ for diffusive boundary conditions.

In summary, we have measured the attenuation of transverse sound in 
$^3$He taking
advantage of the acoustic Faraday effect to determine absolute values. We found
an  anomalous contribution to the attenuation which cannot be accounted for in
terms of collective modes or quasiparticle scattering in the bulk. We suggest
that scattering of transverse sound with surface Andreev bound states
   is the most likely mechanism.  A crossover in the frequency
dependence of the attenuation corresponds to the theoretical value of the upper
limit of the midgap in the surface density of states of
$\Delta^{\star}/\Delta = 0.7$.

We acknowledge support from the National Science Foundation, DMR-0703656 and
thank C.A. Collett, W.J. Gannon and S. Sasaki for useful discussions.

\end{document}